\documentclass[%
 ,twocolumn%
 ,secnumarabic%
,amssymb,nobibnotes, prl, aps]{revtex4}
\usepackage{bm,epsf}%
\newcommand{\bx}{{\bf x}}
\newcommand{\bv}{{\bf v}}
\newcommand{\ba}{{\bf a}}
\newcommand{\bk}{{\bf k}}
\newcommand{\bq}{{\bf q}}
\newcommand{\bp}{{\bf p}}
\newcommand{\bA}{{\bf A}}
\newcommand{\bB}{{\bf B}}
\newcommand{\eps}{{\epsilon}}
\newcommand{\kxy}{{\kappa_{xy}}}
\newcommand{\sxy}{{\sigma^{spin}_{xy}}}
\newcommand{\sone}{{\sigma_1}}
\newcommand{\stwo}{{\sigma_2}}
\newcommand{\sthree}{{\sigma_3}}
\expandafter\ifx\csname package@font\endcsname\relax\else
 \expandafter\expandafter
 \expandafter\usepackage
 \expandafter\expandafter
 \expandafter{\csname package@font\endcsname}%
\fi

\begin{document} \draft

\title{Thermal Hall Conductivity of High Temperature Superconductors: \\
Quantization and Scaling}%

\author{Oskar Vafek and Zlatko Te\v sanovi\'c}
\email{ovafek@pha.jhu.edu, zbt@pha.jhu.edu}
\affiliation{Department of Physics and Astronomy, Johns Hopkins University,
Baltimore, MD 21218
\\ {\rm(\today)}
}
\begin{abstract}
\medskip
Presented is the theory of thermal Hall conductivity $\kappa_{xy}$ in the vortex state 
of high temperature cuprate superconductors in the clean limit. 
We show that $\lim_{T \to 0} \kappa_{xy}/T$ 
is a {\em staircase} function of $1/B$ with an envelope that scales as 
$1/B$. The relation to the experiments is discussed.
\end{abstract}

\maketitle

Fifteen years after the discovery of high temperature superconductivity
in cuprates \cite{BednorzMuller}, the search for the ``final'' 
theory continues unabated. 
Nevertheless, certain important clues about the elements that such a 
theory must contain appear to be established. One of the most significant 
results of the past few years is that, in the superconducting state, cuprates
seem to be well described by the familiar BCS-type formalism with 
Cooper pairs binding in the d-wave channel \cite{harlingen}. 
This is a far from trivial observation in the materials known to exhibit 
strong correlations. 
Soon after the original discovery, Anderson \cite{anderson1} identified 
one of the key features of the new superconductivity: the materials 
are basically two-dimensional with most of the action relevant
for superconductivity taking place 
in the $CuO_2$ planes. The reduced dimensionality and 
the presence of nodal points on an otherwise gapped Fermi surface 
governs the low temperature properties of 
the high-$T_c$ superconductors (HTS). 
Corroborating evidence 
for the BCS character of the fermionic excitations comes from diverse 
spectroscopic and transport experiments \cite{walker, ding, taillefert1, millis}.

Especially informative probe of the quasiparticle dynamics is the thermal Hall 
conductivity, $\kxy$ \cite{ong}, particularly since phonons, 
which usually contribute to the longitudinal 
thermal conductivity, $\kappa_{xx}$, do not contribute to $\kxy$ by the virtue
of their being electrically neutral. 
In addition, from the theoretical point of view, in the leading ``nodal'' approximation, 
$\kxy$ vanishes \cite{simonlee,jinwu,ashvin} and one must consider curvature terms.
For these reasons a special opportunity emerges to scrutinize, 
both theoretically and 
experimentally, the extent to which the fermionic quasiparticles
in HTS exhibit the BCS-type behavior.

Our starting point is a 2-dimensional 
Bogoliubov-de~Gennes Hamiltonian in the vortex state 
\begin{equation}\label{bdg}
H_{BdG}=\sthree\frac{\left(\bp-\sthree\frac{e}{c}\bA\right)^2}{2m^*}-\sthree\mu+
\sone e^{-i\sthree\phi/2}\Delta(\bp) e^{-i\sthree\phi/2}
\end{equation}
where ${\bf A}({\bf x})$ is the vector
potential associated with the uniform external magnetic field $\bB$,
$\bp$ is a momentum operator, and $\sigma$'s are Pauli matrices 
operating in the Nambu space.
The vortex phase fields satisfy
$\nabla\times\nabla\phi ({\bx}) = 2\pi\hat{z}\sum_i
\delta ({\bx} -{\bx}_i)$ with $\bx_i$ denoting the vortex positions
and $\delta ({\bx} -{\bx}_i) $ being a 2D Dirac delta function. For a $d_{x^2-y^2}$-wave 
superconductor $\Delta(\bp)=\frac{\Delta_0}{p_F^2}(p_x^2-p_y^2)$. 

\begin{figure}[t]
\epsfxsize=4.5cm
\hfil\epsfbox{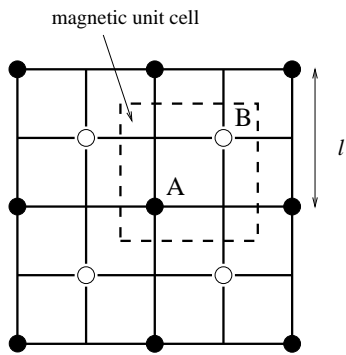}\hfill
\caption{Example of $A$ and $B$ sublattices for the square vortex arrangement.}
\label{transf}
\end{figure}

Following Franz and Tesanovic \cite{ft} (see also \cite{luca,ashvin,vmt}), 
we divide the vortices into two sets $A$ and $B$, positioned at $\{ \bx^A_i\}$ and 
$\{ \bx^B_i\}$ respectively (see Fig. \ref{transf}) and define two phase fields 
$\phi^A(\bx)$ and $\phi^B(\bx)$ such that
$\nabla\times\nabla\phi^{\alpha} ({\bx}) = 2\pi\hat{z}\sum_i\delta ({\bx} -{\bx}^{\alpha}_i),$
 $\alpha=A,B$.
The Hamiltonian (\ref{bdg}) is then transformed into \cite{ft}
\begin{equation}\label{bdgsg}
{\cal H}=U^{\dagger}H_{BdG} U=
\sthree\frac{\left(\bp-\sthree\bv+\ba\right)^2}{2m^*}-\sthree\mu+
\sone \Delta(\bp+\ba) 
\end{equation}
using the unitary operator
\begin{equation}
U=e^{i\chi+i\sthree\Phi}
\end{equation}
where $\chi=\frac{1}{2}(\phi_A-\phi_B)$ and $\Phi=\frac{1}{2}(\phi_A+\phi_B)=\frac{1}{2}\phi$.
In Eq. (\ref{bdgsg}),  $\bv=\frac{1}{2}\nabla\phi-\frac{e}{c}\bA$ 
is the superfluid velocity and $\ba=\frac{1}{2}(\nabla\phi_A-\nabla\phi_B)$ 
is the Berry gauge field
which imposes the condition that a quasiparticle wavepacket 
encircling an $\frac{hc}{2e}$ vortex must pick up an overall minus sign. 

In addition to keeping the wavefunctions single-valued, the advantage of 
this ``bipartite'' singular gauge transformation is that for a periodic
arrangement of vortices (Abrikosov vortex lattice) the resulting
Hamiltonian (\ref{bdgsg}) is itself periodic. 
This is not true of Hamiltonian (\ref{bdg}).
Therefore, the eigenstates of (\ref{bdgsg}) can be expressed through
the familiar Bloch waves $\psi_{n\bk}(\bx)=e^{i\bk\cdot\bx}u_{n\bk}(\bx)$ labeled by one discrete quantum number, 
the band index $n$, and two continuous quantum numbers, the vortex crystal 
momenta $k_x$, $k_y$ that range in the first Brillouin zone.
Note that $u_{n\bk}(\bx)$ is not periodic with the periodicity of the 
vortex lattice, but with the periodicity of the vortex sublattice containing a unit of
{\em electronic} flux $\frac{hc}{e}$ (see Fig. \ref{transf}). 

By extracting the plane wave part of the Bloch states we can define a 
Hamiltonian
\begin{equation}\label{bdgk}
{\cal H}(\bk)=\sthree\frac{\left(\bp+\bk-\sthree\bv+\ba\right)^2}{2m^*}-\sthree\mu+
\sone \Delta(\bp+\bk+\ba)
\end{equation}
where ${\cal H}(\bk)=e^{-i\bk\cdot\bx}{\cal H}e^{i\bk\cdot\bx}$.
We now wish to point out an important feature of ${\cal H}(\bk)$ in the case when the
vortex lattice has an inversion symmetry. 
As shown in Ref.\cite{vmt} an operator
${\cal P}=i\stwo {\cal C}{\cal I}$, where ${\cal C}$ represents charge 
conjugation and ${\cal I}$ takes $\bx$ into $-\bx$, with the following property 
\begin{equation}
{\cal P}^{-1} {\cal H}(\bk) {\cal P}= -{\cal H}(\bk);\;\;
\langle u_{n\bk} |{\cal P}| u_{n\bk} \rangle=0.
\end{equation}
This means that for every energy 
band $\eps_{n\bk}$ there is another $\eps_{m\bk}$ such that $\eps_{n\bk}=-\eps_{m\bk}$, i.e.
the band spectrum has reflection symmetry about the Fermi level and
if a band were to cross the Fermi level ($\eps_{n\bq}=0$ for some $\bq$) 
then there would have to be a twofold degeneracy at $\bq$. However, by the von~Neumann-Wigner 
``non-crossing'' theorem \cite{noncrossing}, the degeneracy unrelated to symmetry
cannot happen for a time reversal breaking Hamiltonian which depends
only on two continuous parameters. Thus, the quasiparticle spectrum of the $d_{x^2-y^2}$ 
superconductor in the vortex state is gapped by a (direct) gap $\Delta_m \ll \Delta_0$ 
unless a third parameter is fine tuned, 
e.g. the band crossing at the Fermi 
level can be achieved by fine-tuning the chemical potential $\mu$.

\begin{figure}[t]
\epsfxsize=8.5cm
\hfil\epsfbox{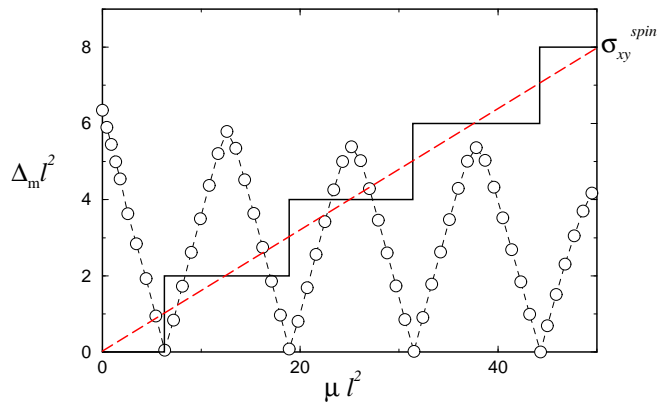}\hfill
\caption{Scaling of the field induced quasiparticle gap $\Delta_m$ (open circles) with
the magnetic field and the chemical potential $\mu$ (measured from the bottom of the
tight-binding band). The quantized values of 
the $\sxy$ are also shown (solid line). 
Magnetic length, $l$, is in units of the tight-binding lattice 
constant $a$ and energies are in units of the hoping energy $t$; the results are
for a square vortex lattice with $\Delta_0=0.1t$. The dashed line is the envelope 
scaling function.}
\label{scalingplot}
\end{figure}

If we rescale all coordinates by the magnetic length
$l=\sqrt{c/eB}$ (see Fig. 1), $\bx \rightarrow l\bx$, then
\begin{equation}\label{operatorscaling}
{\cal H}(\bk,l,\mu,\Delta_0) \rightarrow \frac{1}{l^2}f(l\bk,l^2\mu,\frac{\Delta_0}{p_F^2}),
\end{equation}
where $f$ is a universal (operator) function.
Consequently, the small band gap $\Delta_m$ at the Fermi level, 
which is equal to the minimum of the lowest positive 
eigen-energy of ${\cal H}(\bk,l,\mu)$ with respect to $\bk$, 
is also a universal function that can be written as
\begin{equation}\label{scaling}
\Delta_m(B,\mu,\Delta_0)=B \times F\left(\frac{\mu}{B},\frac{\Delta_0}{p_F^2}\right).
\end{equation}
Thus, in order to determine the dependence of $\Delta_m$ on the magnetic field $B$, 
all we need to know is 
the dependence of $\Delta_m$ on $\mu$ for some fixed B and $\frac{\Delta_0}{p_F^2}$, 
and then use the scaling relation (\ref{scaling}). 
This dependence is determined numerically by diagonalizing 
a tight-binding version \cite{vmt} of the Hamiltonian (\ref{bdgk})
close to the bottom of the tight-binding band where we expect to 
recover the continuum theory (\ref{bdgk}). 
The numerical calculations confirm the scaling  (\ref{scaling}) which gives us
confidence in relating the continuum results to the tight-binding approximation. 
As shown in Fig. \ref{scalingplot}, for a fixed value of 
$\frac{t}{\Delta_0}=\alpha_D=10$, where $t$
is the tight-binding hopping constant,
$\frac{1}{B}\Delta_m(\frac{\mu}{B})$ is a saw-tooth function with a period of approximately $4\pi$.

In Ref. \cite{vmt} it was demonstrated that the gapped quasiparticle 
spectrum leads to the quantization of the spin Hall conductivity
$\sxy$ (see also Ref. \cite{ashvin}), 
which is in turn related to the thermal Hall conductivity, $\kxy$, via 
the Wiedemann-Franz law
\begin{equation}\label{wf}
\lim_{T \to 0}\frac{\kxy}{T}=\frac{4\pi^2}{3}\left(\frac{k_B}{\hbar}\right)^2\sxy(B).
\end{equation}
Because the quantization of $\sxy$ is topological in origin \cite{vmt}, 
any continuous change in a parameter of the Hamiltonian results in 
the same quantized value of $\sxy$ unless the quasiparticle gap collapses. 
Therefore, for a fixed Dirac anisotropy $\alpha_D$, by virtue of Eq. (\ref{operatorscaling}),
the dependence of $\sxy$ on $B$ and $\mu$ can be expressed entirely as a
dependence of $\sxy$ on $\frac{1}{B}\Delta_m(B,\mu)$, i.e. 
\begin{equation}
\sxy(B,\mu)=G\left(\frac{\mu}{B} \right),
\end{equation}
where $G$ is a universal function of its argument that in addition 
depends on the geometry of the vortex lattice. 

For a fixed (large) value of B the numerical computation of $\sxy$ can be readily 
accomplished.  The results for a square vortex lattice and $\alpha_D=10$
are displayed in Fig. \ref{scalingplot} where it is seen that
$\sxy$ is a stair-case function of $\frac{\mu}{B}$ which 
starts at zero at empty filling and then jumps by a multiple of $2\frac{\hbar}{8\pi}$. 
The band crossings, which are responsible for the changes in $\sxy$,  
occur at symmetry points in the Brillouin zone, 
and by the symmetry of the square vortex lattice, the steps 
are bound to come in even integers \cite{notsu2}. 

Thus we arrive at the main result of this Letter shown in
Fig. \ref{scalingplot}: the envelope of the $\sxy(B)$ scales as  $B^{-1}$ and
by Eq. (\ref{wf}) 
\begin{equation}
\lim_{T \to 0} \left(\frac{\kxy}{T}\right) \propto \frac{1}{B}
\end{equation} 
This expression is reminiscent of the electrical Hall conductivity in the 2D electron gas. 

\begin{figure}[t]
\epsfxsize=7.5cm
\hfil\epsfbox{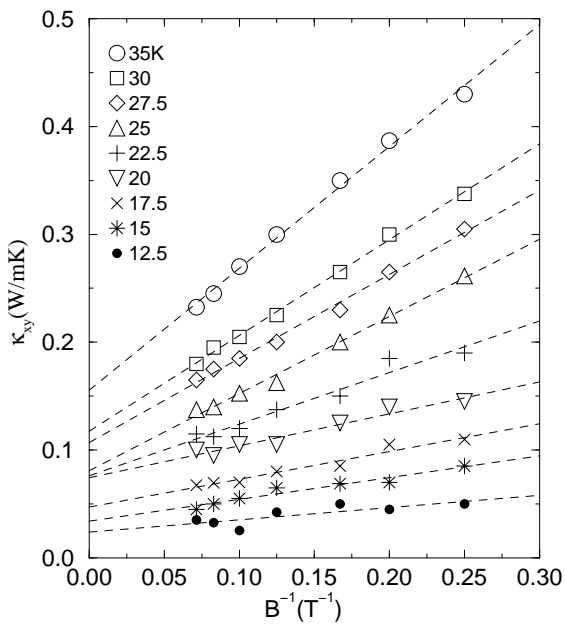}\hfill
\caption{Thermal Hall conductivity data of Zhang et. al. \cite{ong} for $B>4T$ plotted 
vs. $B^{-1}$.}
\label{ongsdata}
\end{figure}

Recently, measurements of $\kappa_{xy}$ were conducted by Ong's group \cite{ong}
on YBCO samples with a very long mean free path.
These experiments were carried out
over a wide range of magnetic fields
(up to $\sim 14T$) and at temperatures from $T \sim 12.5K$ to above the
superconducting transition
$T_c \sim 90K$. Unfortunately, the experiments
are resolution limited below
$12.5K$ as signal becomes too weak. 
In Fig. \ref{ongsdata} we re-analyze the data of Ref. \cite{ong}. 
It is seen that for the magnetic fields above $4T$,  $\kxy\propto C+B^{-1}$ which is, apart 
from the finite intercept, in agreement with our theory.
At fields $B<4T$, $\kxy$ decreases with decreasing $B$ and vanishes at $B=0$ \cite{ong}. 
At low magnetic fields $\frac{1}{B}$ divergence of $\kxy$ is cut-off by disorder  i.e. 
$\kxy=\frac{\rho_{xy}}{\rho_{xx}^2+\rho_{xy}^2}$ with non-vanishing 
longitudinal quasiparticle thermal resistivity $\rho_{xx}$. Even in the presence of weak disorder, 
we still expect $\rho_{xy}\propto B$ and $\rho_{xx}$ to be approximately field 
independent. Therefore, the transverse thermal conductivity in the presence of disorder should vanish 
in the limit of vanishing $B$. 
We must admit that the origin of the residual thermal conductivity 
at $\frac{1}{B}=0$, evident in Fig. \ref{ongsdata}, 
remains mysterious to us and we can only 
speculate that it is not of quasiparticle origin. 

In summary, based on a general analysis of the full {\em non-linearized} 
Bogoliubov-de~Gennes equations,
we argued that magnetic field $B$ will induce a small gap $\Delta_m$
in the quasiparticle spectrum which is a non-monotonic function of the 
magnetic field. Rather, $\Delta_m$ vanishes at some special values of
$\mu/B$ which depend on the details of the vortex lattice and the Dirac cone 
anisotropy (Fig. \ref{scalingplot}). 
The dependence of $\kappa_{xy}$ on $B$ is entirely
due to this non-monotonic behavior of $\Delta_m$, since $\kappa_{xy}/T$ is
a topological quantity and therefore 
can change only at the band crossings \cite{vmt}.  
The topological quantization of $\kappa_{xy}$ was also
discussed in the Ref. \cite{ashvin}.
However, we wish to point out a significant 
qualitative difference between the  
results presented here and those of Ref. \cite{ashvin}:
as we argued before $\kappa_{xy}$ is a quantized staircase function of $1/B$ 
with an {\em envelope}  that itself 
scales as $1/B$, while Ref. \cite{ashvin} states that $\kappa_{xy}/T$ 
can take on only one of the three allowed values $0, \pm 2$
and is otherwise field-independent. Consequently, in a
typical experimental situation, 
the explicit values for $\kappa_{xy}$ obtained from
these two works will differ significantly. 
Still, there is an agreement on the fact that
since the quasiparticle spectrum is gapped $\kappa_{xy}$ does 
not obey simple Simon-Lee scaling, at least when the temperature 
is comparable to the quasiparticle gap. 

We thank Marcel Franz, Ashot Melikyan and Ady Stern 
for comments and discussions. This work was supported in part by NSF 
grant DMR00-94981.

\vspace{-0.5cm}


\begin{references}
\bibitem{notsu2} The trivial double degeneracy of each band due to spin 
is ignored in this calculation.
\end{references}
\end{document}